\documentclass[aps,
groupedaddress,showpacs,nofootinbib,prl,twocolumn]{revtex4}
\usepackage{amssymb,amsmath}
\usepackage{graphics}
\usepackage[english]{babel}
\usepackage{graphicx}
\usepackage{dcolumn}
\usepackage{bm}

\def\fsz{\tiny}
\def\Sigmap{\Sigma
}
\def\Deltap{\Delta
}
\def\rhoO{{\rho_{\sbf 0}}}

\def\rhou{{\rho_{\sbf u}}}
\def\rhosu{{\rho_{\ssbf u}}}

\def\rhotu{{\rho_{\ssbf u}}}
\def\hatTc{{{\hat T}_c^0}}
\def\hatTcO{{{\hat T}_c^0}}

\def\aut#1{#1}

\def\ins#1{{\it #1}}
\def\sdag{\dagger}

\def\meq#1{}
\def\meq#1{}

\def\hc{{\rm h.c.}}

\def\ins#1{}

\def\comment#1{}
\newcommand{\lfrac}[2]{#1/#2}

\def\dst{\displaystyle}
\def\mn#1{\marginpar[]{\scriptsize#1}}
\def\mn#1{}

\def\IncludeEpsImg#1#2#3#4{\renewcommand{\epsfsize}[2]{#3##1}{\epsfbox{#4}}}

\def \xx{\kappa}
\def \GD{\delta}

\usepackage{ulem}
\usepackage{cancel}
\usepackage{color}
\def\rr#1{\textcolor{red}{#1}}

\def\mn#1{\marginpar[\tiny{\rr{#1}}]{\tiny{\rr{#1}}}}

\def\rr#1{}

\usepackage{hyperref}

\def\comment#1{}
\def\IncludeEpsImg#1#2#3#4{\renewcommand{\epsfsize}[2]{#3##1}{\epsfbox{#4}}}

\newcommand{\be}{\begin{equation}}\newcommand{\ee}{\end{equation}}
\newcommand{\bea}{\begin{eqnarray}}\newcommand{\eea}{\end{eqnarray}}
\newcommand{\beaa}{\begin{eqnarray}}\newcommand{\eeaa}{\end{eqnarray}}
\newcommand{\ba}{\begin{array}}\newcommand{\ea}{\end{array}}
\newcommand{\bit}{\begin{itemize}}\newcommand{\eit}{\end{itemize}}
\newcommand{\ben}{\begin{enumerate}}\newcommand{\een}{\end{enumerate}}

 \newcommand{\ssbf}[1]{\mbox{\tiny\bf{#1}}}
 \newcommand{\sbf}[1]{\mbox{\scriptsize\bf{#1}}}

\def\lfrac#1#2{#1/#2}

									\def\be{\begin{equation}}
\def\ee{\end{equation}}				\def\bea{\begin{eqnarray}}				\def\eea{\end{eqnarray}}
\def\bear{\begin{array}}				\def\eear{\end{array}}

												\def\x5{x^{5}}

\begin{document}

\title{Strong-Coupling Bose-Einstein Condensation
}

\author{Hagen Kleinert}
\email{h.k@fu-berlin.de}


\affiliation{Institut f{\"u}r Theoretische Physik, Freie Universit\"at Berlin, 14195 Berlin, Germany}
\affiliation{ICRANeT Piazzale della Repubblica, 10 -65122, Pescara, Italy}


\vspace{2mm}

\begin{abstract}
We extend the theory of Bose-Einstein
condensation from Bogoliubov's weak-coupling 
regime to arbirarily strong couplings.
\end{abstract}

\pacs{98.80.Cq, 98.80. Hw, 04.20.Jb, 04.50+h}

\maketitle


{\bf 1}. For $\phi^4$-theory in $D<4$ Euclidean dimensions
with O($N$)-symmetry, 
a powerful  
strong-coupling theory has been developed in 1998 \cite{strc3}.
It has been carried to 7th order in perturbation theory
in $D=3$ \cite{strc7}, and to 5th order in $D=4-\epsilon$ dimensions \cite{strcep}.
The theory is an extension of a variational approach 
to path integrals 
that was once set by R.P. Feynman and collaborator
in 1989 \cite{FKL}. The extension to high orders
is descibed  
in the textbook
\cite{PI}. 
It is called
{\it Variational Perturbation Theory\/} (VPT).
Originally, the theory was designed 
to convert only the divergent perturbation expansions 
of 
quantum mechanics into exponentially 
fast convergent 
 exressions \cite{JKL}.
In the papers \cite{strc3,strc7,strcep}, it was extended 
from quantum mechanics to  $\phi^4$-theory
with its anomalaus dimensions and produced 
all critical exponents.
This is called quantum field theoretic
VPT.
That theory is explained 
in the textbook \cite{KS} and a recent review  
 \cite{WST}.

Surprisingly, this successful theory has not
yet been applied 
to the presently so popular phenomena 
 of Bose-Einstein 
condensation.
These have so far mainly been focused
\cite{REWS}
on
the semiclassical
treatments
using the good-old Gross-Pitaevskii
equations,  or to 
the weak-coupling 
theory proposed many years ago by Bogoliubov
\cite{BOG}. 
This is somewhat surprising 
since the subject is under 
intense study by many authors.
So far, only 
the 
shift of the critical temperature
has been calculated to high orders \cite{SCT}.
There are only a few exceptions. For instance,
a simple extension of Bogoliubov's thepry to strong couplings was proposed in \cite{ATT}
and pursued further in \cite{ATT2}. But that had an
unpleasant feature that it needed two different chemical potentials
to maintain the long-wavelength properties 
of Nambu-Goldstone excitations required by 
the spontaneously broken U(1)-symmetry 
in the condenste. For this reason it remained widely unnoticed.
Another notable exception is the 
theory in \cite{STOOF}
which came closest to our approach, since it wa also based on 
a variational optimization of the energy.
But by following Bogoliubov in identifying $a_0$ as $\sqrt{\rho_0}$
from the outset, 
they ran into the notorious problem 
of violating the Nambu-Goldstone theorem.
Another approach
that comes close to ours is found in  the paper \cite{STOOF2}.
 Here the main difference lies in the popular use 
the Hubbard-Stratonovic transformation  (HST)
to introduce a fluctuating {\it collective pair field\/}
\cite{CQF}. But, as
pointed out in \cite{hqt} and re-emphasized 
in \cite{HSTP}, 
this makes it impossible 
to  calculate higher-order corrections \cite{HSTP}.

The rules for  applying VPT
to nonrelativistic 
quantum field theories 
in 3+1 dimensions have been specified 
some time ago \cite{SIHF}.
In this note we want to 
show how 
derive from them, to lowest order, the properties
of the Bose-Einstein condensation 
at arbitrarily strong couplings. 

It must be mentioned that in the literature, 
there have been many attempts to treat 
the strong-coupling regime
of various field theories 
for models
with a large number of identical
field components (the so-called large-$N$-models). 
This has first been done for the  so-called spherical 
model \cite{SPHRM}, later the Gross-Neveu 
model
\cite{GN}, and O($N$)-symmetric $\varphi^2$-models \cite{CJP}.
In all these applications, the leading large-$N$ limit has 
been easily solved with the help of the HST 
trick of introducing 
a fluctuating 
field variable
\cite{HST,EPP}
for some dominant 
collective 
phenomenon (Collective Quantum Field Theory \cite{CQF}).
This approach has, however, 
the above-discussed problems 
of going to higher orders 
\cite{HSTP}, which are absent here.

{\bf 2}.
The Hamiltonian of the boson gas has a free term
\begin{eqnarray}
 H_{0} 
\equiv\sum_{\sbf p} a^\sdag_{{\sbf p}}  (\varepsilon_{\sbf p}-\mu)
a_{{\sbf p}}
=\sum_{\sbf p} a^\sdag_{{\sbf p}}  \xi_{\sbf p} a_{{\sbf p}}
,
\label{@SPEns0}\end{eqnarray}
where
$ \varepsilon_{\sbf p}\equiv \lfrac{{\bf p}^2}{2M }$
are the single-particle energies
and  $\xi_{{\sbf p}}
\equiv
 \varepsilon_{\sbf p}-\mu$
the relevant energies 
in a  grand-canonical
ensemble. 
As usual, 
$ a^\sdag_{{\sbf p}} $
and
$ a_{{\sbf p}} $
are creation and annihilation operators
defined by the canonical equal-time  
commutators of the local fields
$\psi({\bf x})=\sum_{{{\sbf p}}}
e^{i{\sbf p}{\sbf x}/\hbar } a_{{\sbf p}}$.
The local interaction 
is
\begin{eqnarray}
 H_{\rm int} =
 \frac{g}{2V} \sum _{{\sbf p},{\sbf p}',
{\sbf q}}
      a^\sdag _{{\sbf p}+{\sbf q} } a^\sdag _{{\sbf p}'-
      {\sbf q}}
       a_{{\sbf p}'} a_{{\sbf p}}.
\label{5.5x}\end{eqnarray}
Instead of following Bogoliubov
in treating the ${\bf p}=0$ modes of the operators $ a_{{\sbf p}}$ 
 classically and identifying 
with the square-root of the condensate density
$\rho_0$,
we introduce  the field  expectation 
$\langle \psi\rangle\equiv \sqrt{V\Sigma_0/g}$
as a {\it variational parameter\/},
and rewrite
$ H_{\rm int}$ as
$ H_{\rm int} ^0=(\lfrac{V}{2g})\Sigma_0^2$ plus 
\begin{eqnarray}
   H_{\rm int} '\!\!=\!\!   
          \frac12
\!\sum _{{\sbf p}\neq {\sbf 0}}\!\!\left[ 2\Sigma_0\!
\left(
a^\sdag_{\sbf p} a_{\sbf p}\!  \!+
\!a^\sdag _{-\sbf p} a_{-\sbf p}\right)
     \!  \!   +\!\Sigma_0\!\!
\left(
           a_{\sbf p}^\sdag  a_{-{\sbf p}}^\sdag
\!+\! \hc\!\right)\! \right] \! ,
\label{@4.newham}\end{eqnarray}
plus a fluctuation Hamiltonian $ H_{\rm int} ''$, which looks like
(\ref{5.5x}), except that the sum contains 
only nonzero-momentum modes.
Now we proceed according to the rules of VPT \cite{HSTP}
and introduce dummy variational parameter 
$\Sigmap $ and $\Deltap$ 
via an auxiliary Hamiltonian
\begin{eqnarray} &&\!\!\!\!\!\!\!\!\!\!\!\!\!\!\!\!
\bar H_{\rm trial}\!=\!-\frac12\!
 \sum _{{\sbf p} \neq0}\!
\left[\Sigmap\!
\left(\!
a^\sdag _{{\sbf p}  }    a_{{\sbf p}}
\!+\!a_{-{\sbf p} }^\sdag a_{-{\sbf p} }    \!\right)\!
 \!  +\!  \Deltap
  a_{- {\sbf p}}
a_{{\sbf p}}  \!+\! \hc
      \right]\!,
\label{@4Fham1BE}\end{eqnarray}
leading a harmonic Hamiltonian
\begin{eqnarray}
\!\! \!\!\!\!\!\!\!\!\!\!\!\!\! H_0'
\!&\equiv&\!
-V\frac{\mu}g \Sigma_0\!+\!
\frac V{2g}\Sigma_{\sbf 0}^2\!
+\!\!
 \sum _{{\sbf p}\neq {\sbf 0}}  \!\!
           \left(  \varepsilon _{{\sbf p}}-\mu+2\Sigma_0  \right)\! a_{\sbf p}^\sdag  a_{\sbf p}
\!   \nonumber \\&+&    \!     \frac{1}{2}\Sigma_0 \!\sum _{{\sbf p}\neq {\sbf 0}}
          \!   \left( a^\sdag_{\sbf p} a_{-{\sbf p}}^\sdag
\! + \!\hc\right)\!+\bar H_{\rm trial} ,
\label{@4.2.76bxZ2}\end{eqnarray}
for which we have to calculate 
the energy prder by order in perturation
theory considering 
\begin{eqnarray}
 H_{\rm int}^{\rm var}=
 H_{\rm int} ''-\bar H_{\rm trial} .
\label{@INteR}\end{eqnarray}
as the interaction Hamiltonian.
The zeroth-order variational 
energy is $W_0=\langle H_0'\rangle $, and the  
lowest-order correction 
comes from the expectation value
$\Delta_1W= \langle H_{\rm int}^{\rm var}\rangle$.
If the energy is calculate to all orders 
in $ H_{\rm int}^{\rm var}$ the result will be independent 
of the variational parameters
$\Sigma_0$, $\Sigma$, and $\Delta$, but the energy 
to any {\it finite\/} order {\it will\/} depend on it.
The optimal values of the parameters are found by optimization 
(usually extremization), and the results converge exponentially fast
as a function of the order \cite{PI,KS,WST}.

A Bogoliubov transformation
with as yet undetermined 
coefficients $  u_{\sbf p}$, $
            v_{\sbf p}$
constrained by the condition 
 $  u_{\sbf p}^2-  v_{\sbf p}^2=1$,
produces a ground state with
vacuum expectation values
$
\langle a_{{\sbf p} }\hspace{-3pt}^\sdag
  a_{{\sbf p}}	 \rangle=
v^ 2_{{\sbf p}}$ and
$
\langle a_{{\sbf p} }
 a_{-{\sbf p} } \rangle=
u_{{\sbf p}}v_{{\sbf p}}$,
so that 
\begin{eqnarray}&&
\!\!\!\!\!\!\!\!\!\!\!\!\!\!\!\!\!\!\!\!\!\!W_0 \!=\!
-V\frac{\mu}g \Sigma_0+\frac V{2g}\Sigma_{\sbf 0}^2
 \nonumber \\&&\!\!\!\!\!\!\!\!\!\!\!\!\!\!\!\!\!\!\!
 +  \sum _{{\sbf p}\neq {\sbf 0}}
  \{\left[ \varepsilon_{{\sbf p}}-\mu-\Sigmap + 2\Sigma_0\right] v^2
   _{\sbf p} + 
({\Sigma_0}-\Deltap)
            u_{\sbf p}
            v_{\sbf p}
\}
.  \label{@4.7BO4} 
\end{eqnarray}

\comment{
Following the rules of VPT in  \cite{SIHF,HSTP},
this energy is the zeroth-order variational energy $W_0$. It has to be improved with the help of an ordinary
perturbation expansion  
 in powers of the modified interaction
Hamiltonian (\ref{@INteR}).
Denoting the result
to any given order $N$  by $W_N$,
this has to be 
 extremized in 
all variational parameters and produces an energy 
$E_N$ 
that approches the exact energy with exponentially increasing accuracy
(like $e^{-c N}$).
}

The first-order variational 
energy $W_1$
contains, in addition,  
the expectation value 
$\langle  H_{\rm int} ^{\rm var}\rangle$.
Of this, the first part,
$\Delta_{(1,0)}W=\langle  H_{\rm int} ''\rangle$,
is found immediately with the  
help of the standard commutation rules
as
a sum 
of three pair terms
\begin{eqnarray} \!\!\!\!\!\!\!\!\!\!
\langle a_{{\sbf p} +{\sbf q}\hspace{1pt}}^\sdag   a_{{\sbf p}'- {\sbf q}\hspace{1pt}}^\sdag
          a_{{\sbf p}'}  a_{{\sbf p} }
\rangle     \!\!\!   &   =&\!\!\!
\langle
a_{{\sbf p} +{\sbf q}\hspace{1pt}}^\sdag
a_{{\sbf p}'- {\sbf q}\hspace{1pt}}^\sdag
\rangle
\langle
 a_{{\sbf p}'}  a_{{\sbf p} }
	\rangle
       \nonumber \\
&&\hspace{-9em}+
\langle
a_{{\sbf p} +{\sbf q}\hspace{1pt}}^\sdag
  a_{{\sbf p} }
\rangle
\langle
a_{{\sbf p}'- {\sbf q}\hspace{1pt}}^\sdag
 a_{{\sbf p}'}
\rangle  +
\langle
a_{{\sbf p} +{\sbf q}\hspace{1pt}}^\sdag
 a_{{\sbf p}'}
	 \rangle
\langle
a_{{\sbf p}'- {\sbf q}\hspace{1pt}}^\sdag
  a_{{\sbf p} }
\rangle.
\label{@4exlue4p}\end{eqnarray}
so that 
\begin{eqnarray}\!\!
\Delta_{(1,0)}W\!=\!\langle  H''_{\rm int} \rangle\!=\! \frac{g}{2V}\!\!
 \sum _{{\sbf p},{\sbf p}'\neq {\sbf 0}}\!
\left(2v^ 2_{{\sbf p}}v^ 2_{{\sbf p}'}+
u_{{\sbf p}}v_{{\sbf p}}u_{{\sbf p}'}v_{{\sbf p}'}
\right).
\label{5.5xpE2}\end{eqnarray}
The second part
$\langle - \bar H_{\rm trial}
 \rangle$  adds to this the expectation value
\begin{eqnarray}
\Delta_{(1,1)} W= 
 \sum _{{\sbf p}\neq {\sbf 0}}\!
\left(\Sigma v^ 2_{{\sbf p}}+
\Delta u_{{\sbf p}}v_{{\sbf p}}
\right).
\label{@DELTASi}\end{eqnarray}

In order to fix the average total number of particles $N$, we
differentiate $W_1\equiv
W_0+\Delta_{(1,0)} W+\Delta_{(1,1)} W$
 with respect 
to $-\mu$ and set the result equal to $N$ 
to find
 the density $\rho=N/V$  as
\begin{eqnarray}
\rho=\frac{\Sigma_0}g+\sum_{\sbf p\neq0}v^2_{\sbf p}.
\label{@NormL}\end{eqnarray}
The momentum  sum  is the density of
particles {\it outside\/} the condensate, 
the {\it uncondensed density}
 \begin{equation}
\rhou=\sum_{{\sbf p}\neq0}\langle
 a^\sdag_{{\sbf p}}  a_{{\sbf p}}
  \rangle=
\frac1V\sum_{{\sbf p}}v_{\sbf p}^2
\comment{=
\frac{1}{2}\int \frac{d^3{p}}{(2\pi\hbar )^3}\left(\frac{
\varepsilon_{\sbf p}+ \bar\Sigma  }{{\cal E}_{\sbf p}}-1\right),
}
\label{@4NUMNP}
\end{equation}
implying that 
$\Sigma_0/g$ is the condensate density
$\rhoO$:
\begin{eqnarray}
\frac{\Sigma_0}g=\rhoO=\rho-\rhou.
\label{@Sdensi}\end{eqnarray}

Now we extremize $W_1$ with respect to the variational parameter 
$\Sigma_0$ 
which yields the equation
\begin{eqnarray}\!\!\!\!\!
\frac{{\mu} -{\Sigma_0}}g\!=\!\!
\sum _{{\sbf p}\neq {\sbf 0}}
(    2  
 v^2_{\sbf p} \!+\!     u_{\sbf p}
            v_{\sbf p})
\!=\!
2\rhou\!+\!\!
\sum _{{\sbf p}\neq {\sbf 0}}
   u_{\sbf p}
            v_{\sbf p}
=2\rhou\!+\!\delta
.
\label{@VARsig0}\end{eqnarray}

We are now  able to fix the size of the Bogoliubov
coefficients
$ u_{\sbf p}$ and
$ v_{\sbf p}$. The original way of doing this 
is algebraic, based on
the elimination of the off-diagonal elements 
of the transformed Hamiltonian operator. 
In the framework of our variational approach it is 
more natural to use the equivalent 
procedure of
{\it extremizing\/} the energy expectation 
 $W_0$
with respect to $u_{{\sbf p}}$ 
and  $v_{{\sbf p}}$ under the constraint
$u^2_{\sbf p}-v^2_{\sbf p}=1$, 
so that $\partial u_{\sbf p}/\partial  {v_{\sbf p}}=v_{\sbf
  p}/u_{\sbf p}$.
\comment{The interaction 
energies 
$\Delta_{(1,0)}W+
\Delta_{(1,1)}W
$ are then  treated
perturbatively.}
Varying $W_0$,
we obtain 
for each nonzero momentum the equation
\begin{eqnarray}~~&&\!\!\!\!\!\!\!\!\!\!\!\!\!\!\!\!\!\!\!\!
 2 \big(\varepsilon_{\sbf p}\!-\mu\! +\!2\Sigma_0-\Sigmap
\big )v_{\sbf p}
\!+\!\big(\Sigma_0\!-\!\Deltap
\big)\!
\left(
u_{\sbf p}
\!+\!v^2_{\sbf p}/
u_{\sbf p}
\right)
 =0.
\label{@@INTeW1}\end{eqnarray}
In order to solve this 
we introduce
the constant 
\begin{eqnarray}
\bar \Sigma\!\equiv\!
\!-\mu\!+\!2\Sigma_0\!-\!\Sigmap
=\!-\mu\!+\!2g(\rho-\rhou)
-\Sigmap
,
\label{@CoNti0}\end{eqnarray}
the right-hand side emerging after using
 (\ref{@4NUMNP}) and
 (\ref{@Sdensi}).

We further introduce the constant 
\begin{eqnarray}
\bar \Delta\equiv 
\Sigma_0-\Deltap
\equiv
r\bar \Sigma
.
\label{@CoNti01}\end{eqnarray}
\comment{
We furthermore 
realize that due to 
the extremization of $W_1$ in $\Sigma_0$,
which led to Eq.~(\ref{@VARsig0}),
we see that
the coefficient
of the second line is related to that of the first line by
\begin{eqnarray}
-\mu+2\Sigma_0+
\frac{g}{V}
 \sum_{{\sbf p}'}
v^2_{{\sbf p}'}=
\Sigma_0+
\frac{g}{V}
 \sum_{{\sbf p}'\neq0}
u_{{\sbf p}'}\,
v_{\sbf p'}.
\label{@@@INTep}\end{eqnarray}
}
Then 
we rewrite (\ref{@@INTeW1})
in the simple form
\begin{eqnarray}~~
 2\!  \left(\varepsilon_{\sbf p}  +\bar\Sigma\right)\!
 v_{\sbf p}+r\bar\Sigma
\left(
u_{\sbf p}
+v^2_{\sbf p}/
u_{\sbf p}
\right)
 =0,
\label{@@INTeW}\end{eqnarray}
which is solved for all ${\bf p}$ by
the Bogoliubov transformation coefficients
\begin{eqnarray}\!\!\!
 u^2_{\sbf p} =
 \frac{1}{2}\left(1+\frac{\varepsilon_{{\sbf p}}
        + \bar\Sigma   }{{\cal E}_{\sbf p}}
\right),~
 v^2_{\sbf p} =-
 \frac{1}{2}\left(1-\frac{\varepsilon_{{\sbf p}}
        + \bar\Sigma }{{\cal E}_{\sbf p}}
\right),
\label{@4.81V'}\end{eqnarray}
with
$
u_{{\sbf p}}v_{{\sbf p}}=\lfrac{\bar\Delta }{2{\cal E}_{\sbf p}}$,
and the   
quasiparticle energies 
\begin{eqnarray}
{\cal E}_{{\sbf p}}=  \sqrt{ \left(  \varepsilon_{{\sbf p}}+\bar\Sigma \right) ^2 -r^2\bar\Sigma^2},
\label{@QPEnw'}\end{eqnarray}

Having determined 
the Bogoliubov coefficients,
we can calculate the above momentum sums
in Eqs.~(\ref{@4NUMNP}) and (\ref{@VARsig0}).
We begin with 
the uncondensed particle density
(\ref{@4NUMNP}). Inserting (\ref{@4.81V'}),
it becomes 
\begin{equation}
\rhou=\frac1V\sum_{{\sbf p}}v_{\sbf p}^2
=\frac{1}{2}\int \frac{d^3{p}}{(2\pi\hbar )^3}\left(\frac{
\varepsilon_{\sbf p}+ \bar\Sigma  }{{\cal E}_{\sbf p}}-1\right)
.
\label{@4NUMNPA}
\end{equation}
The integral is easily done
if we 
set $|{\bf p}|\equiv \hbar k_{\bar\Sigma}\,\kappa  $
with
$k_{\bar\Sigma}=\lfrac{ \sqrt{2M\bar\Sigma } }{\hbar },
$ 
so that we find 
\begin{eqnarray} \!\!\!\!
\rhou=
{k_{\bar\Sigma}^3}
{I_{\rhosu}^{(r)}}/{4\pi^2}
,
\label{@4.ksubgpZ}\end{eqnarray}
where
\begin{eqnarray}\!\!\!\!\!\!\!
\!\!\!
I_{\rhosu}^{(r)}\!\!\!&\equiv&\!\!\!\!\int_0^\infty\!\!
d\xx\,\xx^2\!\left(\!\frac{\xx^2\!+\!1}{\sqrt{(\xx^2\!+\!1)^2-r^2}}\!-\!1\!\right)\!=\!\frac{\sqrt{2}}3f_{\rhosu}^{(r)}.
\label{@zweitep}\end{eqnarray}
with $f_{\rhosu}^{(1)}=1$.
The second momentum sum in 
Eq.~(\ref{@VARsig0})
reads, after inserting (\ref{@4.81V'}),
\begin{equation}\!\!
\GD\equiv
\sum_{{\sbf p}\neq0}\langle
 a_{{\sbf p}}  a_{{\sbf p}}
  \rangle=
\frac
1V\sum_{{\sbf p\neq0}}
u_{\sbf p}v_{\sbf p}=
-r\frac{\bar\Sigma}{2}\int \frac{d^3{p}}{(2\pi\hbar )^3}\frac{
1 }{{\cal E}_{\sbf p}}.
\label{@4NUMNPi}\end{equation}
In contrast to (\ref{@4NUMNPA}), this is a divergent quantity.
As a consequence of the renormalizability of the theory, the divergence can be removed
  by absorbing it 
into the inverse coupling constant of the model
defined by 
\begin{eqnarray}
\frac1{g_R}\equiv \frac1g-
\frac{1}{V}\sum _{{\sbf p}\neq {\sbf 0}} \frac{1}{{2\varepsilon_{\sbf p}}}
= \frac1g-
\int \frac{d^3{p}}{(2\pi)^3}\frac{1}{2\varepsilon_{\sbf p}},
\label{@REngR}\end{eqnarray}
The renormalized coupling is finite and measurable 
in two-body scattering
as an $s$-wave scattering length: 
$g_R=2\pi \hbar^2 a_s/M$.
Thus we 
introduce the finite  renormalized quantity
\begin{equation}
\GD_R=\frac1V\sum_{{\sbf p}}u_{\sbf p}v_{\sbf p}
=-
\frac{r\bar\Sigma }{2}\int \frac{d^3{p}}{(2\pi\hbar )^3}
\left(
\frac{
1 }{{\cal E}_{\sbf p}}
-\frac{
1 }{{\varepsilon}_{\sbf p}}
\right),
\label{@4NUMNPii}\end{equation}
and write
$
\GD=\GD_R+\GD_{\rm div},
$ where the divergence is the momentum  sum
\begin{equation}
\GD_{\rm div}\equiv
-\frac{r \bar\Sigma}V\sum_{{\sbf p}}\frac{1}{2\varepsilon_{\sbf p}}
=-
\frac{r\bar\Sigma }{2}\int \frac{d^3{p}}{(2\pi\hbar )^3}
\frac{
1 }{{\varepsilon}_{\sbf p}}
\label{@4NUMNPi4}\end{equation}
If we denote this  by $-\bar\Sigma/Vv$,
 we have 
\begin{eqnarray}
\GD=
\GD_R+\GD_{\rm div}=\GD_R-\frac{r\bar\Sigma}{Vv}
.
\label{@PSgapp}\end{eqnarray}
Inserting 
this
together with
(\ref{@4NUMNP}) 
 into (\ref{@VARsig0}),
we find
\begin{eqnarray}
\frac{{\mu} -{\Sigma_0}}g=2\rhou+\GD_R+\GD_{\rm div}.
\label{@EXTrat}\end{eqnarray}
Recalling (\ref{@Sdensi}), this implies
\begin{eqnarray}
\frac{\mu}g =
\rhoO+2\rhou+\GD_R+\GD_{\rm div}=\rho+\rhou+\GD_R+\GD_{\rm div}
.
\label{@MurelaT'}\end{eqnarray}

If we evaluate
the momentum sum
(\ref{@4NUMNPii}) in the same way as
(\ref{@4NUMNPA}), it yields
\begin{eqnarray}
\delta_R=k_{\bar\Sigma}^3
\lfrac{I_\delta^{(r)}}{4\pi^2} ,
\label{@KGDEFP}
\label{@GaPB1}\end{eqnarray}
where $I_\delta
$ is given by the integral 
\begin{eqnarray}\!\!\!\!\!\!\!\!\!\!
I_\delta^{(r)}\!\!\!&\equiv&\!\!\!-r \int_0^\infty\!\!
d\xx\,\xx^2\!\left(\!\frac 1{\sqrt{(\xx^2\!+\!1)^2\!-\!r^2}}\!-\!\frac1{\xx^2}\!\right)\!=\!
\sqrt{2}f_\delta^{(r)},
\label{@GaPB}\end{eqnarray}
with $f_{\delta}^{(1)}=1$.

%
\comment{
Note that (\ref{@GaPB1}) looks like a bosonic version of the superconducting gap equation. 
Indeed, 
we may express   
the renormalized coupling in terms of the $s$-wave scattering length via
$g_R=4\pi \hbar^2 a_s/M$,
and using (\ref{@KGDEFP}),
we can rewrite the
(\ref{@GaPB1})
as
\begin{eqnarray}
-\frac{1}{a_sk_{\bar \Sigma}}=\frac2 \pi I_1.
\label{@QgEqp}\end{eqnarray}
We
 we shall 
refer to this
as
the {\it quasigap equation\/} of the Bose gas.
\ins{quasigap equation}%
\ins{equation,quasigap}%
There is, however, and 
important 
difference with respect a 
proper
 gap
equation. 
The present quasigap parameter $\bar\Sigma$
does {\it not\/} determine an energy gap, but the energy {\it slope\/} in the 
{\it gapless\/}  energy spectrum 
(\ref{@QPEnw'}).
As such, it determines the small-${\bf p}$ slope in the
new quasiparticle energy 
 (\ref{@QPEnw'}):
\begin{eqnarray}
{\cal E}_{{\sbf p}}=  \sqrt{ 2\bar\Sigma \varepsilon_{{\sbf p}}+
  \varepsilon^2_{{\sbf p}}}\approx \sqrt{ \bar\Sigma/M}|{\bf p}|
+{\cal O}({\bf p}^2).
, 
\label{@QPEnw'}\end{eqnarray}
and thereby the velocity of second sound 
as
$
c\equiv
 \sqrt{\lfrac{\bar\Sigma }{M}}$,
}

We continue the discussion with Eq.~(\ref{@CoNti0}), which we rewrite 
using (\ref{@MurelaT'}) as
\begin{eqnarray}
\frac{\bar\Sigma}g
\!=\!
\rho\!-3\!
\rhou\!
-\!\frac{\Sigma}g
-\!\GD_R\!+\!\frac{\bar \Sigma}{Vv}
.
\label{@256n}\end{eqnarray}
As before in Eqs.~(\ref{@4NUMNPi}),
(\ref{@REngR}), 
and (\ref{@4NUMNPii}), the last, divergent  term can be absorbed into the 
first by renormalizing the coupling constant, so that we obtain
\begin{eqnarray}
\frac{\bar\Sigma}{g_R}=\rho-
3\rhou-\GD_R
\!
-\!\frac{\Sigma}g
.
\label{@256n'}\end{eqnarray}

Finally, we calculate the total variational  energy
$W_1$. 
Inserting
the Bogoliubov coefficients 
(\ref{@4.81V'})
into $W_0$ of Eq.~(\ref{@4.7BO4}) 
and adding 
the action energies
$\Delta_{(1,0)}W+
\Delta_{(1,1)}W
$
of
and (\ref{5.5xpE2}) and (\ref{@DELTASi}),
we 
have 
\begin{eqnarray}\!\!\!
\!\!\!\!\!\!\!\!\!\!\!\!\!\!\!W_1 &\!=\!&
-\frac V g{\mu}\Sigma_0+\frac {V}{2g}\Sigma_0^2
 +w_0(\bar \Sigma,r)
-\frac{r^2\bar \Sigma^2}{4Vv}
\nonumber \\&+&\Delta_{(1,0)}W
+\Delta_{(1,1)}W
,  \label{@4.7BO5A} 
\end{eqnarray}
where $w_0(\bar \Sigma)$ is the convergent momentum sum 
\begin{eqnarray}
w_0(\bar \Sigma,r)\equiv
\frac12  \sum _{{\sbf p}\neq {\sbf 0}}
  \left\{\left[ {\cal E}_{{\sbf p}}-\varepsilon_{{\sbf
        p}}-\bar\Sigma+\frac{r^2\bar \Sigma^2}{2\varepsilon_{{\sbf
        p}}}\right]
\right\}.
\label{@W0Energy}\end{eqnarray}
This is evaluated as in
 (\ref{@4NUMNPA}) to
\begin{eqnarray}&&
\!\!\!\!\!\!\!\!\!\!\!\!\!\!\!w_0(\bar \Sigma)  =
 V \bar \Sigma k^3_{\bar \Sigma}\lfrac{I_{E}^{(r)}}{4\pi ^2},
  \label{@4.7BO6} 
\end{eqnarray}
where
\begin{eqnarray}
\label{@nullte}\!\!\!\!
I_E^{(r)}\!&\equiv&\!\!\!
\int _0^\infty\!
d\xx\xx^2\left[\sqrt{(\xx^2\!+\!1)^2\!-\!r^2}\!-\!\xx^2\!-\!1\!+\!\frac{r^2}{2\xx^2}\right]\!
\nonumber \\&=&\!\frac{8\sqrt{2}}{15}f_E^{(r)},
\end{eqnarray}
with $f_E^{(1)}=1$.
If we rename  all  $I_E/4\pi^2$ to $\bar I_E$,
the energy $W_1$ becomes
\begin{eqnarray}\!\!\!\!\!\!\!\!\!\!\!\!
\!\!\!\!\!\!\!\!\!\!\!\!\!\!\!\!\!\!\!\!\!\!\!\!\!\!\!\!\!
\!\!\!\!\!\!\!\!\!\!\hspace{-1cm}W_1 \!\!\!
& =&\!\!
-\frac V g{\mu}\Sigma_0+\frac {V}{2g}\Sigma_0^2
+
\frac V2\bar\Sigma
k^3_{\bar \Sigma}\bar I_E^{(r)}
 \label{@4.7BO5X} 
 \\
&+&\!\!
\frac{Vg}2 k^{3} _{\bar \Sigma}(
2 \bar I_{\rhosu}^{(r)}{}^2\!\!+\!\Delta \bar I_\delta^{(r))}
{}^2)
\!+\!
{V} k^{3/2} _{\bar \Sigma}{(\Sigma
 \bar I_{\rhosu}^{(r)}\!\!+\!\Delta \bar I_\delta^{(r)})}
. \nonumber 
\end{eqnarray}
The expression is renormalized 
most simply using
\comment{
The divergent contributions
\begin{eqnarray}
W_1^{\rm div}=
V\frac{\bar\Sigma\Sigma_0}{2v}
- \frac{ \bar \Sigma^2 }{ 4v}
- \frac{ \Sigma\bar \Sigma}{ 4v}+\frac g{8V} \frac{\bar
  \sigma^2}{v^2}-\frac g 2
 \frac{\bar \Sigma\delta_R}v.
\label{@}\end{eqnarray}
have been omitted on the basis of
}dimensional regularization, that allows us to use  Veltman's rule 
\cite{KS} to set $1/v=0$.
\comment{
Extremizing 
(\ref{@4.7BO5}) with respect to $\Sigma$
remembering (\ref{@DeltArel}), we fix
$\Sigma$ as a function of $\bar \Sigma$.
}

We now prepared to  extremize the variational energy
Eq.~(\ref{@4.7BO5X}) with respect 
to $\bar\Sigma$ and $\bar\Delta$.
We insert 
$\Sigma/g\equiv 
\rho-3\rho_{\sbf u}
- \delta_R-\bar \Sigma/g$ from 
(\ref{@256n'}) and 
$\Delta/g\equiv \rho- \rho_{\sbf u}
-\bar \Delta/g$ 
from (\ref{@CoNti01})
and vary $W_1$
in 
$\delta\Sigma$ and $\delta\Delta$. 
This yields the equations
\begin{eqnarray}
&&\!\!\!\!\!\!\!\!(\lfrac{\bar\Sigma}g -\rho^{(-)})S_{11}+
(\lfrac{\bar\Delta}g -\rho^{(+)})S_{12}=0,\\
&&\!\!\!\!\!\!\!\!(\lfrac{\bar\Sigma}g -\rho^{(-)})S_{21}+
(\lfrac{\bar\Delta}g -\rho^{(+)})S_{22}=0,
\label{@}\end{eqnarray}
where
$\rhou^{(\pm)}\equiv\rho- \rhou\pm\delta_R$ and
\comment{
\begin{eqnarray}
&&
\!\!\!\!\!\!\!\!
S_{11}=- \frac {\bar \Delta^2} V\sum_{\sbf p\neq0} \frac{1}{2{\cal E}_{\sbf p}^3},~~~
S_{12}= \frac {\bar \Delta} V\sum_{\sbf p\neq0} \frac{
\varepsilon_{\sbf p}+\bar \Sigma
}{2{\cal E}_{\sbf p}^3},~~~
\\
&&
\!\!\!\!\!\!\!\!
S_{21}= \frac{\bar \Delta} V\sum_{\sbf p\neq0} \frac{
\varepsilon_{\sbf p}+\bar \Sigma
}{2{\cal E}_{\sbf p}^3},~~~
S_{22}=\frac 1 V\sum_{\sbf p\neq0} \frac{
(\varepsilon_{\sbf p}+\bar \Sigma)^2
}{2{\cal E}_{\sbf p}^3},
\label{@}\end{eqnarray} 
}
\begin{eqnarray}
&&
\!\!\!\!\!\!\!\!
\!\!\!\!\!\!\!\!
S_{11}=\bar I_{\rhosu}^{(r)} \lfrac{\partial k^3_{\bar \Sigma}}{\partial\bar \Sigma},~~~
S_{12}=(\lfrac{k^3_{\bar \Sigma}}{\bar \Sigma})\lfrac{\partial\bar I_{\rhosu}^{(r)}}{\partial r},~
\\
&&
\!\!\!\!\!\!\!\!
\!\!\!\!\!\!\!\!
S_{21}=\bar I_{\rhosu}^{(r)}\lfrac{\partial k^3_{\bar \Sigma}}{\partial\bar \Sigma},~~~
S_{22}=(\lfrac{k^3_{\bar \Sigma}}{\bar \Sigma})\lfrac{\partial\bar I_\delta^{(r)}}{\partial r}.~
\label{@BothEq}\end{eqnarray} 
These equations are solved 
for $r=1$
with 
\begin{eqnarray}
\frac{\bar \Sigma}g=
\frac{\bar \Delta}g=\rho-\rhou-\delta_R.
\label{@ReSuw}\end{eqnarray}
The reason is simply that 
both 
$\bar I_{\rhosu}^{(r)}$ and
$\bar I_\delta^{(r)}$
behave near $r=1$ like $(1-r)^{3/2}$
so that  derivative ar $r=1$ vanishes
and both equations 
(\ref{@BothEq})
give (\ref{@ReSuw}).

The solution of these equations
is $r=1$,
thus
guaranteeing the Nambu-Goldstone nature of the
quansiparticle energies (\ref{@QPEnw'}).

{\bf 3}.
To extract experimental
consequences it is
useful to re-express all equations 
in a dimensionless 
form by
introducing the reduced variables 
\begin{eqnarray}
s\equiv\frac{\bar \Sigma}{\varepsilon_a},
\label{@sSIgma}\end{eqnarray}
where
$
\varepsilon_a\equiv\lfrac{\hbar^2}{2Ma^2}$ is the natural  energy 
scale 
of the system.
We also introduce the reduced
$s$-wave scattering length
\begin{eqnarray}
\hat a_s\equiv 8\pi \frac{  a_s}{a},
\label{@RSWSL}\end{eqnarray}
in terms of which the renormalized coupling constant 
is
\begin{eqnarray}
 g_R=\frac{4\pi\hbar ^2}M a_s=8\pi \varepsilon_a a^2a_s= \varepsilon_a
 a^3 \hat a_s ,
\label{@GRnat}\end{eqnarray}
while
\begin{eqnarray}
k_{\bar \Sigma}\!=\!\frac{\sqrt s}{a},
~~~~  \frac{\bar \Sigma}{g_R}=\frac s{8\pi a^2a_s}
=\frac s{a^3\hat a_s},
\label{@kREpl}\end{eqnarray}
and the second-sound velocity reads
\begin{eqnarray}
c=\sqrt{\frac s 2}\,v_a,~~~~~v_a\equiv\frac
{p_a}M\equiv\frac{\hbar}{aM}.
\label{SECSv}
\end{eqnarray}
\comment{
Then we may write 
(\ref{@GaPB1})
as
$ \hat \delta\equiv \delta_R/\rho=s^{3/2}
 \bar I_\delta$,
and
(\ref{@4NUMNPA}) as
$\lfrac{\rhou}{\rho}=s^{3/2}
 \bar I_\rhosu$, so that Eq.~(\ref{@256n'})
reads
\begin{eqnarray}
\frac{s}{\hat a_s}=1-s^{3/2}(
 \bar I_\rhosu+\bar I_\delta)-{s_\Sigma}
,
\label{@256n''}\end{eqnarray}
$\sigma_\Sigma\equiv 1-\hat \rho_{\sbf u}-\hat \delta-s/\hat a_s$ from 
(\ref{@256n'}).
}

Let us also define the reduced quantity
$\hat \delta\equiv s^{3/2}\bar I_{\delta}$
and
\begin{eqnarray}
\hat \rho_{\sbf u}\equiv s^{3/2}\bar I_{\rhosu}.
\label{ehrouvs}\end{eqnarray}
In terms of these 
we calulate the reduced variational energy
$ w_1\equiv W_1/ N\varepsilon_a$ 
from Eq.~(\ref{@4.7BO5X}) for $r=1$:
\begin{eqnarray}
w_1\! \! &=&\!
-{\hat a_s}
(1\!+\!\hat \rho_{\rm u}\!+\!\hat\delta)(1\!-\!\hat \rho_{\rm u})
\!+\!\frac{{\hat a_s}} 2
(1\!-\!\hat \rho_{\rm u})^2\!+\!\frac{s^{5/2}}2\bar I_E\nonumber \\
&+&
\frac{\hat a_s}2(2\hat \rho_{\rm u}^2+
\hat {\delta}^2)
+
\hat a_s(\sigma_\Sigma\hat \rho_{\rm u}+
\sigma_\Delta \hat \delta)
,
  \label{@4.7BXX} 
\end{eqnarray}
where 
$\sigma_\Sigma\equiv 1-3\hat \rho_{\sbf u}-\hat \delta-s/\hat a_s$ from 
(\ref{@256n}) and 
$\sigma_\Delta\equiv 1-\hat \rho_{\sbf u}
-s/\hat a_s$ 
from (\ref{@CoNti01}). 
Inserting here
$\hat\rho_{\sbf u}$ and 
$\hat \delta$, and going 
from the grand-canonical to the 
true proper energies by adding $\mu N$ 
 to $W_1$ forming
$W^{\rm e}=W_1+\mu V\rho$, we obtain
the reduced energy
\begin{eqnarray}
w^{\rm e}_1=\frac{\hat a_s}2-\frac{4\sqrt{2}}{15\pi^2}s^{3/2}
+\frac{\sqrt{2}}{3\pi^2}\hat a_ss^{3/2}
+\frac{1}{72\pi^4}\hat a_s^{4}.
\label{@}\end{eqnarray}
The relation between $s$ and $\hat a_s$
is from (\ref{@ReSuw})
\begin{eqnarray}
\frac{s}{\hat a_s}=1-s^{3/2}(\bar I_\rhosu+\bar I_\delta).
\label{@sversaa}\end{eqnarray}
leading to expansion
\begin{eqnarray}\!\!\!\!
w^{\rm e}_1=\frac{\hat a_s}2+\frac{\sqrt{2}}{15\pi^2}
\hat a_s^{5/2}+\frac{1}{72\pi^4}\hat a_s^4+\dots~.
\label{@}\end{eqnarray}
Note that in the strong-coupling limit,
$s\rightarrow s^{\rm sc}=(3\pi^2/\sqrt{2})^{1/3}$,
the maximal depletion is 
$\hat \rho{}_{\bf u}=1/4$, and the energy behaves 
like $w^{\rm e}_1\rightarrow B+ A\,\hat a_s $, with 
$B=-4 \times 2^{1/12}/(5 \times 3^{1/6}\pi^{1/3})\approx-0.48$, and
$A=1/2+1/24\sqrt{2}\pi^2+2 ^{1/4}/\sqrt{3}\pi\approx 0.72$.
The sound velocity at infinite coupling is 
$c=\sqrt{ s^{\rm sc}/2}v_a$.

Let us now study the temperature dependence of our results.
For this 
we
 introduce the temperature-dependent 
version of the integral
(\ref{@zweitep})
at $r=1$, where we omit the trivial superscript,
to find
\begin{eqnarray} \!\!\!\!
\rhou(t)= \! k_{\bar\Sigma}^3
\lfrac{ I_{\rhosu}(t)}{4 \pi^2},~~~~ k_{\bar\Sigma}=
\lfrac{ \sqrt{2M\bar\Sigma } }{\hbar }=\lfrac{\sqrt{s}}a.
\label{@4.ksubgp1}\end{eqnarray}
where $I_{\rhosu}(t)$ is defined by the integral
\comment{
\begin{eqnarray}\!\!\!\!\!\!\!
\Delta h_2(t)=\frac{1}{s^{3/2}}
\frac3{\sqrt{2}}
\int_0^\inft\!
d\xx\xx^2\frac{\xx^2\!+\!s}{\sqrt{(\xx^2+s)^2-s^2}}
\frac {2}{e^{\sqrt{(\xx^2+s)^2-s^2}/\hat T}-1} 
\!
,
\label{@TdepH2}\end{eqnarray}
}
\begin{eqnarray}\!\!\!\!\!
I_\rhosu(t)\!\!&\equiv&\!\!\!\!
 \int_0^\infty\!\!
d\xx\xx^2\!\left[\frac{\xx^2\!+\!1}{\sqrt{(\xx^2\!+\!1)^2\!-\!1}}
c_t(\kappa)-1\right]
\!,
\label{@zweitep'}
\end{eqnarray}
where
$c_t(\kappa)\equiv
\coth
\left(\!\lfrac{\sqrt{(\xx^2+1)^2-1}}{2t}\!\right),
$
and $t$ is the reduced temperature
\begin{eqnarray}
t\equiv k_BT /\varepsilon_{ \bar\Sigma},~~~
\varepsilon_{ \bar\Sigma}\equiv\sqrt{2M\bar \Sigma}.
\label{@RedT1}\end{eqnarray}
To find the $s$ 
at any temperature we need 
Eq.~(\ref{@KGDEFP})
for $T\neq 0$, where it reads
\begin{eqnarray} \!\!\!\!
\delta_R(t)= \! k_{\bar\Sigma}^3
\lfrac{I_{\delta}(t)}{4\pi^2},
\label{@KGDEFP1}\end{eqnarray}
with 
\begin{eqnarray}
\!\!\!\!\!
I_{\delta}(t)\!\!&\equiv&\!\!-
 \int_0^\infty\!\!
d\xx\xx^2 \!\!\left[\!
\frac{1}{\sqrt{(\xx^2\!+\!1)^2\!-\!1}}
c_t(\kappa)
-\!
\frac1{\xx^2}
\!\right]
\!.
\label{@zweitepD}\end{eqnarray}
to find
\begin{eqnarray} \!\!\!\!
\GD_R\!=
 k_{\bar\Sigma}^3
\frac{I_\delta(t)}{4\pi^2}.
\label{@4.ksubgp1D}\end{eqnarray}
\comment{t
Here we write alternatively
\begin{eqnarray}
h_1(t)=-\frac1{\sqrt{2}}
 \int_0^\infty
d\xx\,\xx^2\left[
\frac{1}{\sqrt{(\xx^2+1)^2-1}}
\left(1+\frac {2}{e^{\sqrt{(\xx^2+1)^2-1}/t}-1}\right)-\frac1{\xx^2}\right].
\label{@4.ksuXbgpH}\end{eqnarray}
so that
\begin{eqnarray}\!\!\!\!\!\!
h_1(t)\equiv 1+\Delta h_1(t)=1
-\frac1{\sqrt{2}}\int_0^\infty
d\xx\,\xx^2\frac{1}{\sqrt{(\xx^2+1)^2-1}}
\frac {2}{e^{\sqrt{(\xx^2+1)^2-1}/t}-1} 
.
\label{@TH1int}\end{eqnarray}
}
For the temperature dependence of the energy in 
Eqs.~
(\ref{@W0Energy}),
and (\ref{@4.7BO6}),
we calculate  
\begin{eqnarray}
I_E(t)
\equiv\!\!\!
\int _0^\infty\!
d\xx\xx^2\left[\left(\sqrt{(\xx^2\!+\!1)^2\!-\!1}\!-\!\xx^2\!-\!1\right)
c_t(\kappa)\!+\!\frac1{2\xx^2}\right].
\nonumber\!\!\!\!\!\! \\
\label{@}\end{eqnarray}

\comment{By analogy with (\ref{@4.ksubgp})
and
(\ref{@4.ksubgpm}), we shall rewrite the
depletion equation (\ref{@4.ksubgp1})
as
\begin{eqnarray} \!\!\!\!
\frac{\rhou}\rho = s^{3/2}\frac{I_2}{4\pi^2}h_2(t)=
 s^{3/2}\frac{I_2}{4\pi^2}+\frac{\Delta\rhou}{\rho} , 
\label{@4.ksubgpx1}\end{eqnarray}
with
\begin{eqnarray}
\frac{\Delta\rhou}\rho =s^{3/2}\frac{I_2}{4\pi^2}h_2(t).
\label{@}\end{eqnarray}
}

The phase transition lies at the temperature where $\rhou=\rho$.
Along the transition, we find from Eq.~(\ref{@sversaa})
the relation between $s$ and $\hat a_s$
\begin{eqnarray}
   - \lfrac{s}{\hat a_s}=s^{3/2}{\bar I_\delta}.
\label{SAPHSt}
\end{eqnarray}
For weak couplings, all expressions 
can be calculated analytically.
The calculation is somewhat subtle
since the small $s$ -region of the integral cannot simply 
be obtained by expanding the integrand in powers of $s$,
 the first  correction going like $\sqrt{s}$.
To see this,
we must proceed as in the derivation of the Robinson
expansion of the Bose-Einstein integral function  \cite{remaRo}, writing 
$\lfrac{\rhou}\rho=s^{3/2}I_2/4\pi^2+\lfrac{\Delta\rhou}\rho$, with
the second term  being 
the integral 
\begin{eqnarray}\!\!\!
\frac{1}{4\pi^2}
\int_0^\infty
d\xx\,\xx^2\frac{\xx^2\!+\!s}{\sqrt{(\xx^2\!+\!s)^2\!-\!s^2}}
\frac {2}{e^{\sqrt{(\xx^2+s)^2-s^2}/\tau \hatTc}-1} 
,
\label{@TdepH2}\end{eqnarray}
where $\hatTcO\equiv T_c^0/T_a=4\pi \zeta(3/2)^{-2/3}$ 
 is the reduced critical
temperature of the free Bose gas,
and
$\tau$ is the ratio $T/T_c^0$.
The integral can be done immediately for $s=0$
and yields the well-known result
\begin{eqnarray}\!\!\!\!\!\!\!
\frac{\Delta\rhou}\rho
\mathop{=}_{s\rightarrow 0}\frac{\Delta\rhou^0}\rho=
\tau^{3/2}
.
\label{@TdepH2'}\end{eqnarray}
For small $s$, there is an additional subtracted  term
\begin{eqnarray}\!\!\!\!\!\!
\hspace{-4em}\frac{\Delta\rhou'}\rho&=&
\frac{1}{4\pi^2}
\int_0^\infty
d\xx\,\xx^2\bigg[\frac{\xx^2+s}{\sqrt{(\xx^2+s)^2-s^2}}
\nonumber \\&\times&
\frac {2}{e^{\sqrt{(\xx^2+s)^2-s^2}/\tau\hatTc}-1}
-
\frac {2}{e^{\xx^2/\tau\hatTc}-1} 
\bigg],
\label{@TdepH2}\end{eqnarray}
The first term 
takes its leading small-$s$ behavior 
from the linear  Nambu-Goldstone
momentum behavior
of second sound, becoming
\begin{eqnarray}\!\!\!\!\!\!\!
\frac{\Delta\rhou'}\rho\approx
\frac{2\tau \hatTc}{4\pi^2}
\int_0^\infty
d\xx\,\xx^2\left(
\frac {\kappa^2+s}{{\xx^4+2s \xx^2}}
-
\frac {1}{\xx^2} 
\right),
\label{@TdepH2'W}\end{eqnarray}
which is equal to
$-({\tau\hatTc}/4{\pi})\sqrt{\lfrac{s}2}$.
Thus we obtain for small $s$
the leading terms
of the uncondensed particle density
\begin{eqnarray} \!\!\!\!
\frac{\rhou}\rho = 
 s^{3/2}\frac{\sqrt{2}}{3\cdot 4\pi^2}
+\tau^{3/2}
-\frac{\tau\hatTc}{4\pi}\sqrt{\frac{s}2}+
\dots
 . 
\label{@4.ksubgpx1}\end{eqnarray}
The last term is dominant for small $s$.
Its negative sign has an interesting effect upon
the phase diagram
observed in earlier publications, that for
small coupling constant, 
the critical temperature {\it increases\/} above the free Bose gas
 value $\hatTc$
to  $\hat T_c=\tau_c\hatTc$
with
\begin{equation}
    \tau_c=1+\frac{2}3\frac{\hatTc}{4\pi}
\sqrt{\frac{s}2}
+{\cal O}(s).
\label{TcCurve}
\end{equation}
A similar  limit square-root limit
appears in the relation
(\ref{@KGDEFP1})
for $\delta_R$
which becomes 
\begin{eqnarray}
\frac{\delta_R}{\rho}= s^{3/2}\frac{I_\delta(t)}{4\pi^2}
=
s^{3/2}\frac{\sqrt{2}}{4\pi^2}
+
\frac{\sqrt{2}}{4\pi^2}s^{3/2}\Delta h_1(t),
\label{@DelatRrel}\end{eqnarray}
where the last term is equal to
\begin{eqnarray}\!\!\!\!\!\!
-\frac{s}{4\pi^2}\!\!\int_0^\infty\!\!
d\xx\,\xx^2\frac{1}{\sqrt{(\xx^2\!+\!s)^2\!-\!s^2}}
\frac {2}{e^{\sqrt{(\xx^2+s)^2-s^2}/\tau\hatTc}\!-\!1} 
.
\label{@TH1intE}\end{eqnarray} 
To lowest order in $\tau$ this 
yields
\begin{eqnarray}-\frac{s}{4\pi^2}
\int_0^\infty
d\xx\,\frac{2\tau\hatTc}{\xx^2+2 s }
 =-\frac{\tau\hatTc}
{4\pi}\sqrt{\frac{s}2}
,
\label{@}\end{eqnarray}
so that  we find
\begin{eqnarray}
\frac{\delta_R}{\rho}= s^{3/2}\frac{I_1(t)}{4\pi}
=
s^{3/2}\frac{\sqrt{2}}{4\pi^2}
-\frac{\tau\hatTc}{4\pi}
\sqrt{\frac{s}2}.
\label{@}\end{eqnarray}
According to 
(\ref{SAPHSt}), this is equal to $-s/\hat a_s$, implying for small 
$s$, where the $s^{1/2} $-term is dominant,
the relation between $s$ and $\hat a_s$ 
 along the phase
transition line
\begin{eqnarray}
\sqrt{\frac{s}2}
\approx
\frac{\hatTc}{8\pi}\,
\hat a_s.
\label{@}\end{eqnarray}
Inserting this into 
(\ref{TcCurve}), we obtain 
\begin{eqnarray}
\tau_c=1+\frac{1}3\frac{({\hatTc})^2}{(4\pi)^{2}}\hat a_s+\dots~,
\label{@PhaSET}\end{eqnarray}
which becomes with $\hatTc=
 [\zeta(3/2)]^{-2/3} 4\pi$:
\begin{eqnarray}
\frac{ T_c}{T_c^0}
=1+\frac{\hat a_s}{3\zeta(3/2)^{4/3}}
+\dots
=1+C\,
\frac{a_s}a+\dots~,
\label{@PhaSET}\end{eqnarray}
where the
constant is 
$C\approx 8\pi /3\pi \zeta(3/2)^{4/3}\approx$
$ 2.33$.
This shows the  initial increase 
of the critical temperature for small repulsion 
between the bosons discussed in \cite{REENTR}.

Numerically, the prefactor $C$
of the linear term  is twice as big
as the value 
$C^{\rm 5\,loop}_{\rm VPT}=0.93\pm 0.13$
predicted from 5-loop variational
perturbation theory in Ref.~\cite{SCT}, and 
83\% larger
than 
the value 
 $C=1.27\pm0.11$
from its extension
to 
seven-loops \cite{KAS}.
It is the same as the value obtained from 
a large-$N$ calculation \cite{Baym}.
 
\begin{figure}[htb]
\vspace{-.6cm}
\hspace{-0cm}
\unitlength1pt
\hspace{-13em}
\begin{picture}(94.55,75)
\input epsf.sty
\def\dst{\displaystyle}
\def\IncludeEpsImg#1#2#3#4{\renewcommand{\epsfsize}[2]{#3##1}{\epsfbox{#4}}}
\put(-5,0){\IncludeEpsImg{94.49mm}{64.13mm}{0.3000}{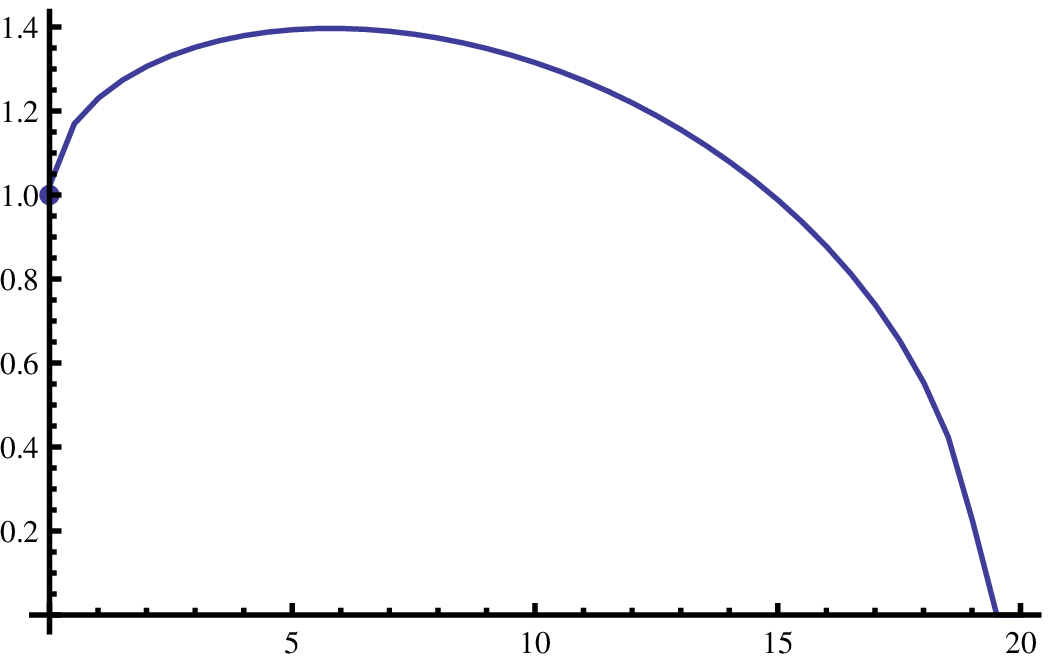}}
\put(112,0){\fsz b)}
\put(-15,0){\fsz a)}
\put(15,37){\fsz $\tau=T/T_c^0$}
\put(25,-5){\fsz $s=\bar \Sigma/\varepsilon_a$}
\put(125,0){\IncludeEpsImg{94.49mm}{64.13mm}{0.3000}{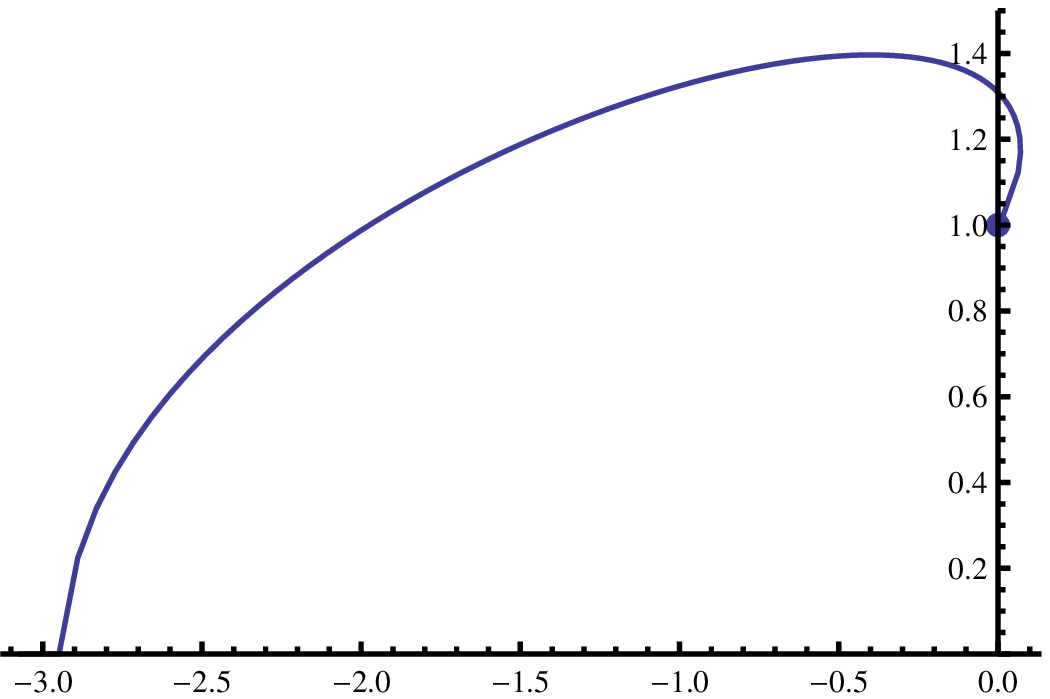}}
\put(160,33){\fsz $\tau=T/T_c^0$}
\put(170,-5){\fsz $\hat a_s$}
\end{picture}
\caption[Transition temperature $T/T_c^0 as
a function of s\equiv \bar \Sigma/\varepsilon_a$
]{Transition temperature $\tau=T/T_c^0     $ as
a function of a) $s\equiv \bar \Sigma/\varepsilon_a$, and b) of
of the reduced $s$-wave scattering length $\hat a_s$.
}
\label{tcvons}\end{figure}

For strong coupling, there is no subtlety and the 
phase transition can be extracted from a numerical 
plot of the location where  
(\ref{ehrouvs}) is equal to unity.
Some plots are shown in 
Fig.~\ref{tcvons}.
The  limit of infinitely strong
coupling 
is found from the vanishing of 
the right-hand side of Eq.~(\ref{@sversaa}).
In particular we can easily calculate the 
temperature dependence 
of the second-sound velocity
$\sqrt{s/2}v_a$ as a function of temperature using
Eq.~(\ref{@sversaa}).

\begin{figure}[t]
\vspace{-.6cm}
\hspace{-0cm}
\unitlength1pt
\hspace{-12em}
\begin{picture}(94.55,75)
\input epsf.sty
\def\dst{\displaystyle}
\def\IncludeEpsImg#1#2#3#4{\renewcommand{\epsfsize}[2]{#3##1}{\epsfbox{#4}}}
\put(110,0){\fsz b)}
\put(-20,0){\fsz a)}
\comment{
\put(118,2){\IncludeEpsImg{94.49mm}{64.13mm}{0.3000}{rhovonaT.eps}}
\put(130,52){\fsz $\rho_{\ssbf u}/\rho$}
\put(163,-2){\fsz $\hat a_s$}
\put(211,42){\fsz $\tau=1$}
\put(211,16.5){\fsz $\tau=0$}
\put(211,21){\fsz $\tau=0.25$}
\put(211,26){\fsz $\tau=0.5$}
\put(211,32){\fsz $\tau=0.75$}
}
\put(118,2){\IncludeEpsImg{94.49mm}{64.13mm}{0.3000}{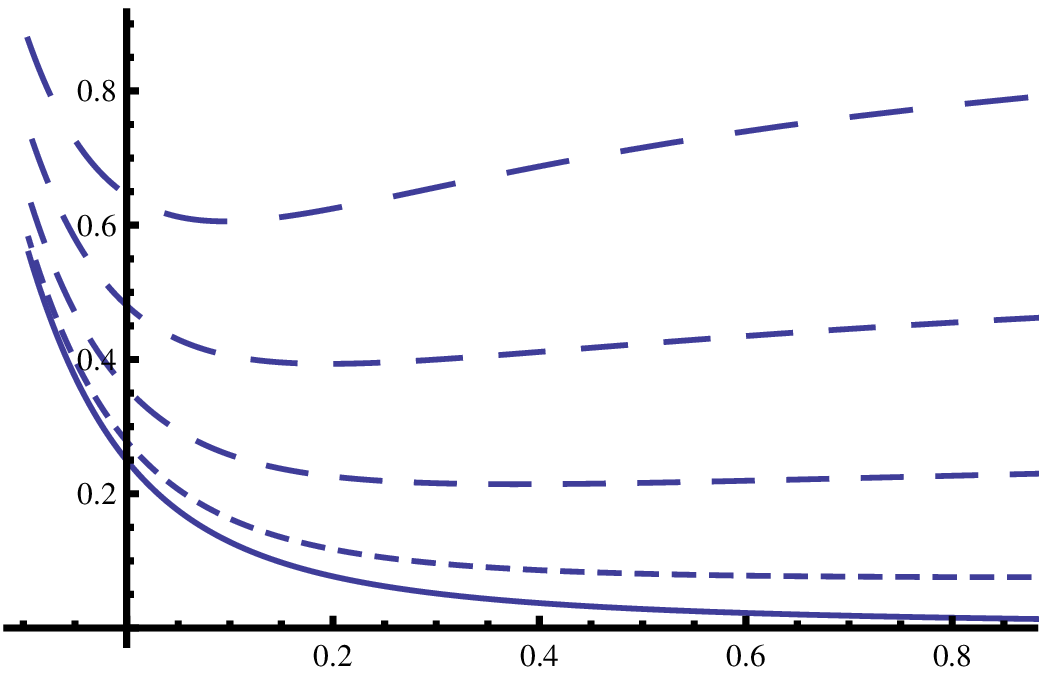}}
\put(145,52){\fsz $\rho_{\ssbf u}/\rho$}
\put(163,-2){\fsz $1/\hat a_s$}
\put(211,52){\fsz $\tau=1$}
\put(211,6){\fsz $\tau=0$}
\put(211,10){\fsz $\tau=0.25$}
\put(211,18){\fsz $\tau=0.5$}
\put(211,32.5){\fsz $\tau=0.75$}

\put(-10,0){\IncludeEpsImg{94.49mm}{64.13mm}{0.3000}{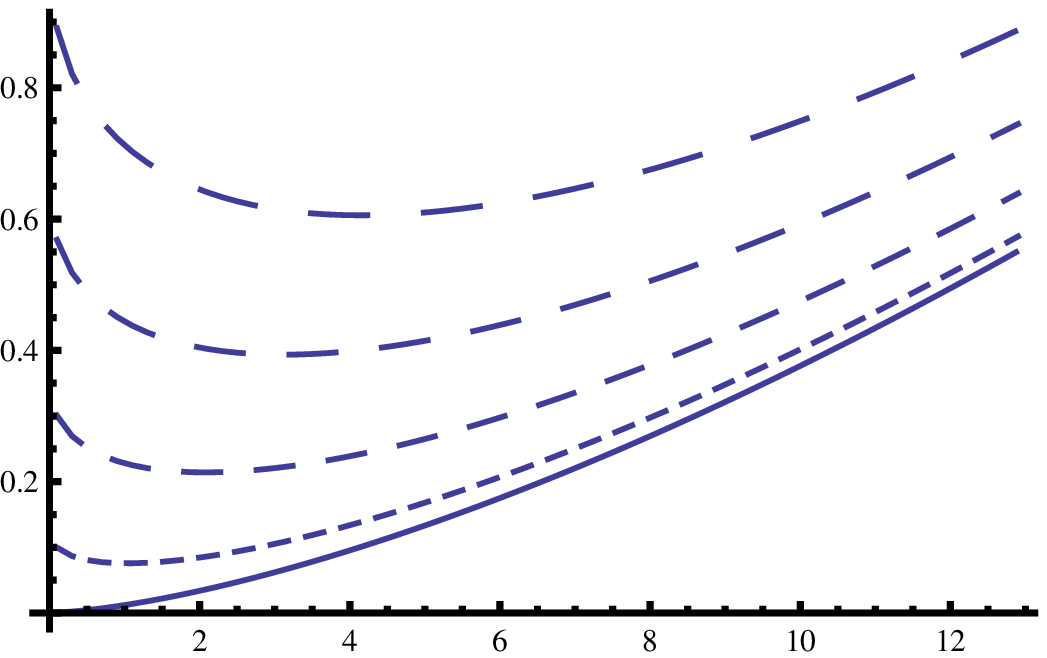}}
\put(16,52){\fsz $\rho_{\ssbf u}/\rho$}
\put(82,55){\fsz $\tau=1$}
\put(82,32.5){\fsz $\tau=0$}
\put(82,37){\fsz $\tau=0.25$}
\put(82,48){\fsz $\tau=0.75$}
\put(82,42){\fsz $\tau=0.5$}
\put(39,-4){\fsz $s$}
\end{picture}
\caption[Uncondensed density  $\rhou$ as a function of
a) the quasigap parameter $s$ and b) the reduced $s$-wave scattering length $\hat
a_s$  at various 
reduced
temperatures $T/T_c^0=\tau=0$, $0.25$, $0.5$
]{a) Uncondensed density  $\rhou$ as a function of
a) the quasigap parameter $s$ and b) the reduced $s$-wave scattering length $\hat
a_s$  at various 
reduced
temperatures $T/T_c^0=\tau=0$, $0.25$, $0.5$.  
Note that the curves 
continue smoothly to negative $1/\hat
a_s$. 
}
\label{thosvaT}\end{figure}

\begin{figure}[t]
\vspace{-.6cm}
\hspace{-0cm}
\unitlength1pt
\hspace{-12em}
\begin{picture}(94.55,75)
\input epsf.sty
\def\dst{\displaystyle}
\def\IncludeEpsImg#1#2#3#4{\renewcommand{\epsfsize}[2]{#3##1}{\epsfbox{#4}}}
\put(112,0){\fsz b)}
\put(-20,0){\fsz a)}
\comment{
\put(-10,0){\IncludeEpsImg{94.49mm}{64.13mm}{0.3000}{crossBEC.eps}}
\put(45,48){\fsz $s=\bar \Sigma/\varepsilon_a$}
\put(20,-5){\fsz $s/\hat a_s$}
\put(80,3){\fsz $\tau=0$}
\put(-3,8){\fsz $\tau=1.5$}
\put(35,8){\fsz $1$}
\put(60,8){\fsz $0.5$}
}

\put(-10,0){\IncludeEpsImg{94.49mm}{64.13mm}{0.3000}{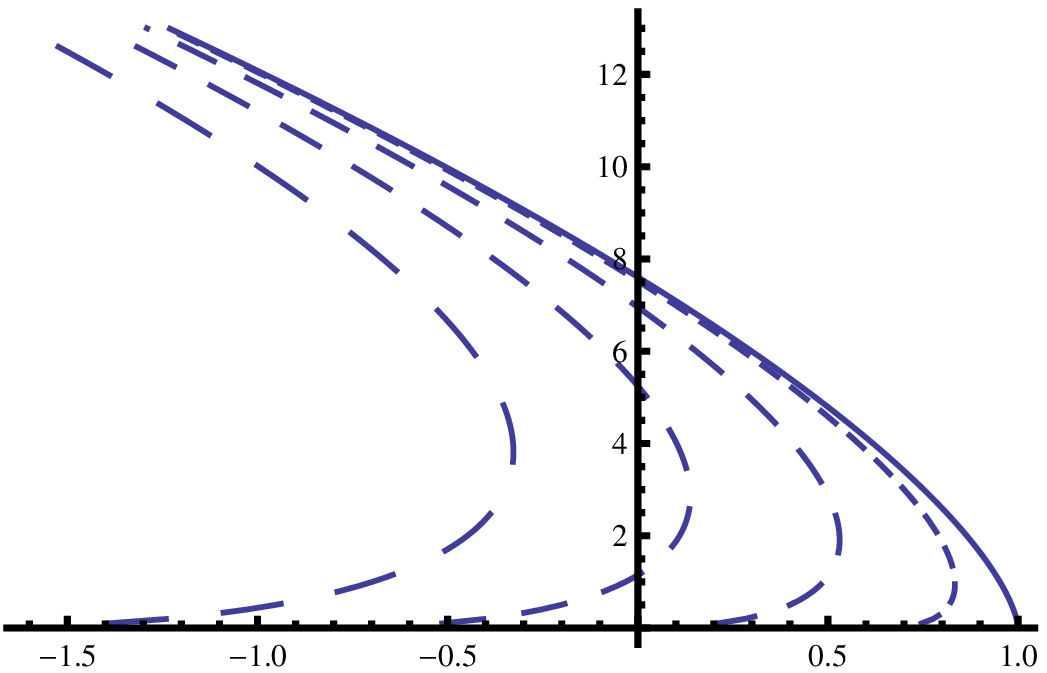}}
\put(55,48){\fsz $s=\bar \Sigma/\varepsilon_a$}
\put(36,-5){\fsz $s/\hat a_s$}
\put(78,7){\fsz $\tau=0$}
\put(30,16){\fsz $2$}
\put(40,13){\fsz $1.5$}
\put(55,10){\fsz $1$}
\put(63,8){\fsz $0.5$}

\comment{
\put(115,0){\IncludeEpsImg{94.49mm}{64.13mm}{0.3000}{rhonofa.eps}}
\put(150,63){\fsz $\rho_{\ssbf u}/\rho$}
\put(186,63){\fsz $\tau=1.5$}
\put(186,8){\fsz $\tau=0.5$}
\put(186,35){\fsz $\tau=1$}
\put(186,-2){\fsz $\tau=0$}
\put(200,20){\fsz $s/\hat a_s$}
}
\put(125,0){\IncludeEpsImg{94.49mm}{64.13mm}{0.3000}{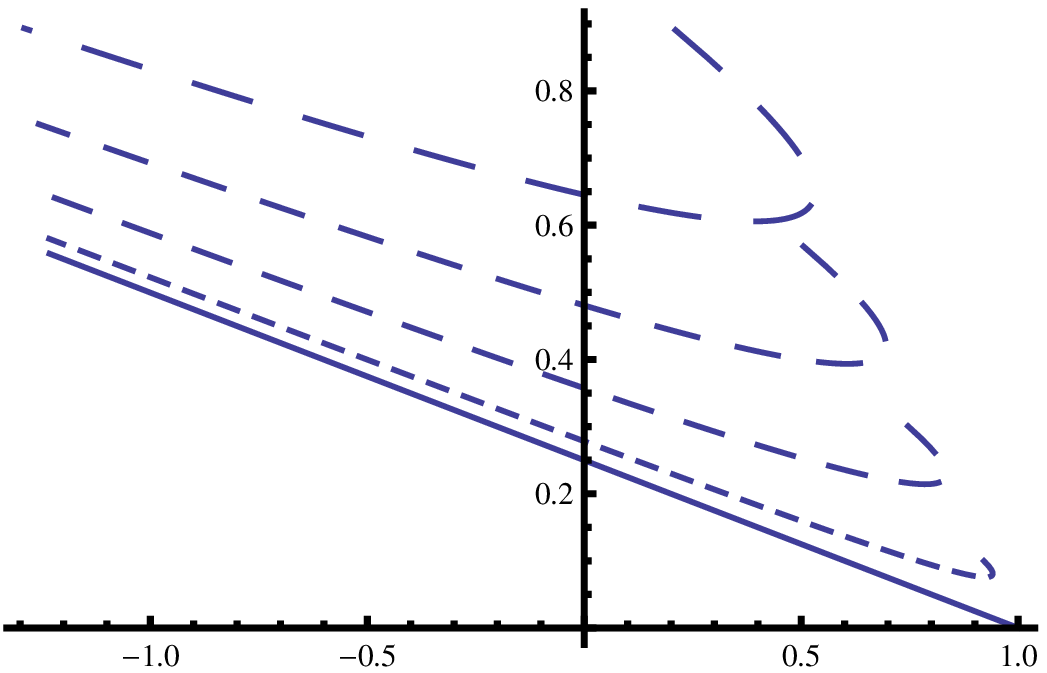}}
\put(150,59){\fsz $\rho_{\ssbf u}/\rho$}
\put(213,10){\fsz $\tau=0.25$}
\put(206,28){\fsz $\tau=0.75$}
\put(211,19){\fsz $\tau=0.5$}
\put(201,40){\fsz $\tau=1$}
\put(216,5){\fsz $\tau=0$}
\put(170,-5){\fsz $s/\hat a_s$}
\end{picture}
\caption[Relation between the reduced $s$-wave scattering length $\hat
a_s$
and the quasigap parameter $s$  determining the second-sound
velocity via $c=\sqrt{s/2}v_a$
]{a) Relation between the reduced $s$-wave scattering length $\hat
a_s$ and quasigap parameter  $s=\bar \Sigma/\varepsilon_a$
determining the second-sound
velocity via $c=\sqrt{s/2}v_a$,
 at various 
reduced
temperatures $\tau=T/T_c^0 =
0$, $0.5$, $1$, $1.5$, and $2$.
b) Density of uncondensed fluid $\rhou$ as a function of $s/\hat a_s$
at reduced 
temperatures $\tau=0$, $0.25$, $0.5$, $0.75$,
and $1$.
}
\label{negas}\end{figure}

The behavior of Eqs.~(\ref{@sversaa})
and (\ref{ehrouvs}) is
shown in Figs.~\ref{thosvaT}
and 
\ref{negas}.
An interesting feature of
 Fig.~\ref{thosvaT}b) and
 Figs.~\ref{negas}
is that the curves continue smoothly
beyond the strong-coupling limit $1/\hat a_s=0$ 
to negative values.
This should be observable
in experiments 
through a Feshbach resonance.

\comment{
An interesting feature of Fig.~\ref{thosvaT}b)
is that the curves continue smoothly
beyond the strong-coupling limit $1/\hat a_s=0$ 
to negative values.
This should be observable
in experiments 
through a Feshbach resonance.
}

{\bf 4}.
All properties 
of the strongly interacting Bose gas determined by the above
theory
can be calculated in the presence of superflow
of velocity $v$, by 
simply generalizing 
 the integrals 
(\ref{@zweitep'}) and
(\ref{@zweitepD})
for $I_\rhosu(t)$
and
$I_\delta(t)$
to 
$I_\rhosu(t,\nu)$
and
$I_\delta(t,\nu)$,
these being defined  by
interchanging in each integrand the terms $(\kappa^2+1)$  
by $(\kappa^2+1+2\kappa\nu/s^{3/2})$,
where $\nu$ is the reduced velocity of the gas $\nu\equiv v/v_a$.
From 
the associated second derivative of the energy 
we can easily find the superfluid density 
$\rho_s$ as a function of the velocity.

There is  no problem
to drive the accuracy to any desired level, with exponentially fast
convergence, as was demonstrated 
by the calulation 
of critical exponents 
in all Euclidean $\varphi^4 $ theories with $N$ components in $D$
dimensions
\cite{KS}.
The procedural rules were explained 
in the paper
\cite{HSTP}. We merely have to calculate 
higher-order diagrams using 
the harmonic Hamiltonian 
(\ref{@4.2.76bxZ2}) as the free theory
that determines the Feynman diagrams, 
and (\ref{@INteR}) as the
 interaction Hamiltonian that determines the vertices.
At any given order, the results are optimized 
in the variational parameters 
$\Sigma_{0},\Sigmap$,
and
$\Deltap$.
The theory is renormalizable,
so that 
all divergencies 
can be absorbed in a redefintion 
of the parameters 
of the orginal action, order by order.
This is the essential advantage of the present
 theory
over any previous 
strong-coupling 
scheme published so far in the literature,
in particular  over those based on
Hubbard-Stratonovic transformations
of the interaction, 
which are applicable only in some large-$N$
limit as explained in \cite{HSTP}, and for which no higher-loop
calculations are renormalizable.

\comment{
Observe that if we were to include
 $\Delta_{(1,1)}W$  into 
the first-order variational energy $W_0+
\Delta_{(1,0)}W$,
the 
variational parameters
$\Sigmap$
and
$\Deltap$
of Eq.~(\ref{@4Fham1BE})
would 
disappear completely.
The procedure applied here, however, 
makes the optimization
with respect to
$\Sigmap$
and
$\Deltap$ essential in producing 
compatibility with the Nambu-Goldstone theorem.
}

Our results can be made much more reliable 
in the $\bar \Sigma\neq 0$ -regime by calculating
the contribution
of the still-missing second two-loop diagram.\footnote{The second diagram 
in  Eq.~(3.741) of the textbook
\cite{PI}. Its contribution would be the $3+1$-dimensional version 
of the last term 
in Eq. (3.767), is essential in the $X\neq0$ phase.
Without this term, the slope of the quantum-mechanical energy as a
function of the coupling constant 
is missed by 25\%, as discussed in the heading of Fig. 5.24.
}

~\\
Acknowledgement:
I am grateful to
Axel Pelster
and
Aristeu Lima
 for interesting discussions, and
to
Flavio Nogueira
and Henk Stoof 
for useful comments.


\begin{thebibliography}{9}


\bibitem{strc3}
H. Kleinert,
Phys. Rev. D 57, 2264 (1998) (cond-mat/9801167) .

\bibitem{strc7}

H. Kleinert, Phys. Rev. D 60, 085001 (1999) (hep-th/9812197).

\bibitem{strcep}

H. Kleinert,
Phys. Lett. B 434, 74 (1998) (cond-mat/9801167).


\bibitem{FKL}
R.P Feynman and H. Kleinert,
Phys. Rev. A 34, 5080 (1986). 



\bibitem{PI}
H. Kleinert, \textit{Path  {I}ntegrals in {Q}uantum {M}echanics,
  {S}tatistics, {P}olymer {P}hysics, and {F}inancial {M}arkets}, 5th
ed., World Scientific, 2009 ({\tt klnrt.de/b5}).


\bibitem{JKL}
W. Janke and H. Kleinert,
Phys. Rev. Lett. 75, 2787 (1995) (quant-ph/9502019 ) 


\bibitem{KS}

H. Kleinert and Schulte-Frohlinde,
{\it Critical Properties of $Phi^4$-Theories},
World Scientific, Singapore 2001
 ({\tt klnrt.de/b8}).

\bibitem{WST}
H.~Kleinert,
EJTP 8, 15 (2011)
{\tt (www.ejtp.com/\linebreak articles/ejtpv8i25p15.pdf)}.

\bibitem{REWS}                                                     

P.W. Courteille, V.S. Bagnato, and V.I. Yukalov, Laser Phys.
     11, 659 (2001͒);
C.J. Pethic and H. Smith, {\it Bose-Einstein Condensation in Dilute
Gases\/}
(Cambridge University Press, Cambridge, 2002);
L. Pitaevskii and S. Stringari, {\it Bose-Einstein Condensation\/}    
    (Clarendon, Oxford, 2003͒);


\bibitem{BOG}
N.N. Bogoliubov,
J. Phys. (Moscow) 11, 23 (1947);
{\it Lectures on Quantum Statistics\/},
Godon and Breach, N.Y., 1979).



\bibitem{SCT}
H. Kleinert,
Mod. Phys. Lett. B 17, 1011 (2003) ({\tt  klnrt.de/320}).
This paper goes up to five loops in VPT.



\bibitem{KAS}
The five-loop calculation of Ref. \cite{SCT}
has been carried to seven loops by
B. Kastening,
Phys. Rev. A 69, 043613 (2004).

\comment{
\bibitem{LOOPSCT}
B. Kastening,
Phys. Rev. A 69, 043613  (2004).
}



\bibitem{ATT}
V. I. Yukalov and H. Kleinert, 
Phys. Rev. A 73, 063612 (2006) 


\bibitem{ATT2}
V. I. Yukalov and
 E. P. Yukalova,
Phys. Rev. A 76, 013602 (2007) .


\bibitem{STOOF}
M. Bijlsma and H.T.C. Stoof, 
Phys. Rev. A 55, 498-512 (1997). 

\bibitem{STOOF2}
A. Koetsier, P. Massignan, R.A. Duine, and H.T.C. Stoof,
 Phys. Rev. A 79, 063609 (͑2009͒).


\bibitem{CQF}

H. Kleinert, {\it Collective Quantum Fields},
     Lectures presented at the First Erice Summer School on
     Low-Temperature Physics, 1977,
     Fortschr. Physik   {26}, 565-671 (1978) ({\tt klnrt.de/55/55.pdf}).


\bibitem{SIHF}
H. Kleinert,
Annals of Physics 266, 135 (1998) 
({\tt  klnrt.de/255}).



\bibitem{SPHRM}
H.E. Stanley,
Phys. Rev. 176, 718–722 (1968) 

\bibitem{GN}
D.J. Gross A. Neveu, Phys. Rev. D 10, 10
(1974).


\bibitem{CJP}
S. Coleman, R. Jackiw, and D. Politzer,
Phys. Rev. D 10, 2491–2499 (1974). 





\bibitem{HST}
R.L. Stratonovich, Sov. Phys. Dokl. {2}, 416 (1958),
J. Hubbard, Phys.\ Rev.\ Letters {3}, 77 (1959);
B. M\"uhlschlegel,
J. Math.\ Phys.\, { 3}, 522 (1962); J. Langer, Phys.\ Rev.\
   {134}, A 553 (1964); T. M. Rice, Phys.\ Rev.\ {140} A 1889
(1965); J. Math.\ Phys.\ {8}, 1581 (1967); A. V. Svidzinskij,
  Teor.\ Mat.\ Fiz.\ {9}, 273 (1971); D. Sherrington, J. Phys.\
 {C4} 401 (1971).
%
\bibitem{EPP}
The first authors to employ such identities were \\
P. T. Mathews,
 A. Salam, Nuovo Cimento {12}, 563 (1954), { 2}, 120 (1955).\\
Applications came with many models in large-$N$ limits
such as nonlinear $\sigma$ models four-Fermi theories.


\bibitem{hqt}
H. Kleinert, {\it
On the Hadronization of Quark Theories},
Lectures presented at the Erice Summer Institute 1976, in
Understanding the Fundamental Constituents of Matter,
Plenum Press, New York, 1978, A. Zichichi ed., pp. 289-390
 ({\tt klnrt.de/53/53.pdf}).



\bibitem{HSTP}
H. Kleinert,
EJTP 8, 57 (2011), {\tt (www.ejtp.com/\linebreak articles/ejtpv8i25p57.pdf)}.


\bibitem{remaRo}
\aut{J.E. Robinson}, Phys. Rev. {83\/}, 678
  (1951).
See also the textbook \cite{PI} on p. 172.



\bibitem{REENTR}

H. Kleinert, S. Schmidt, and A. Pelster,
Phys. Rev. Lett. 93, 160402 (2004);
Ann. Phys. (Leipzig) 14, 214 (2005).
\\
A recent discussion 
and comparison of various data is found in\\
K. Morawetz, M. Männel, and M. Schreiber,
Phys. Rev. B 76, 075116 (2007͒).


\bibitem{BaymZ}

G. Baym, J.-P. Blaizot, J.Zinn-Justin,
Europhys. Lett. 49, 150 (2000). 

\comment{
\bibitem{KAS}
B. Kastening,
Phys.Rev. A 69, 043613 
(2004).
}
\bibitem{Baym}
G. Baym, J.-P. Blaizot, and J. Zinn-Justin, Europhys. Lett. 49, 150
(2000).
\end{thebibliography}
 
\end{document}